\documentclass{article}
\usepackage{spconf,amsmath,graphicx}
\usepackage{multirow}
\usepackage{amssymb}
\usepackage{float}
\usepackage{float}
\usepackage{stfloats}
\usepackage{soul} 
\usepackage{color, xcolor} 

\title{Towards Speaker Age Estimation with Label Distribution Learning}
\name{Shijing Si, Jianzong Wang$^{*}$\footnotemark[1], Junqing Peng and Jing Xiao}
\address{Ping An Technology (Shenzhen) Co., Ltd.}

\begin{document}
%
\maketitle

\renewcommand{\thefootnote}{\fnsymbol{footnote}} 
\footnotetext[1]{Corresponding author: Jianzong Wang, jzwang@188.com}
\begin{abstract}
Existing methods for speaker age estimation usually treat it as a multi-class classification or a regression problem. However, precise
age identification remains a challenge due to label ambiguity, \emph{i.e.}, utterances from adjacent age of the same person are often indistinguishable. 
To address this, we utilize the ambiguous information among the age labels, convert each age label into a discrete label distribution and leverage the label distribution learning (LDL) method to fit the data. For each audio data sample, our method produces a age distribution of its speaker, and on top of the distribution we also perform two other tasks: age prediction and age uncertainty minimization. Therefore, our method
naturally combines the age classification and regression approaches, which enhances the robustness of our method.
We conduct experiments on the public NIST SRE08-10 dataset and a real-world dataset, which exhibit that our method outperforms baseline methods by a relatively large margin, yielding a 10\% reduction in terms of mean absolute error (MAE) on a real-world dataset.

\end{abstract}
\section{Introduction}
\label{sec:intro}

Accurate estimation of speaker age in speech is integral to various speech technology applications \cite{tawara2021age}, for instance, user-profiling, targeted marketing, or personalized call-routing.
For systems operated with voice, which are
increasingly popular nowadays, the age information can be
helpful to adapt such systems to the user giving a more natural human-machine interaction \cite{schuller2013paralinguistics}. Call centers can also benefit from these
systems in order to classify speakers in age categories or to perform user-profiling.

Many researchers have attempted to perform speaker age estimation. According to \cite{Sadjadi_2016}, existing methods can be roughly classified into two types: 1.) feature based methods \cite{tawara2020improving,kalluri2019deep} which focus on extracting robust features from input audio to predict the age using standard classification/regression algorithms; 
and 2.) back-end based methods \cite{Ghahremani_2018,kitagishi2020speaker} where the goal is to either develop or identify a classification/regression algorithm that can effectively estimate the age information from standard speech representations such as the mel-frequency cepstral coefficients (MFCC). For the feature based methods, i-vector \cite{Bahari_2014}, d-vector \cite{variani2014deep} and x-vector \cite{snyder2018x}, which map variable-length utterances to fixed-dimensional embeddings, are developed. Fedorova \cite{fedorova2015exploring} used i-vectors combined with a separate deep neural networks (DNNs) back-end for regression.
For the back-end methods, Minematsu \cite{minematsu2002automatic} utilized the MFCC features with Gaussian mixture models (GMM) for binary age classification. 
\cite{Ghahremani_2018} developed an end-to-end deep neural networks (DNNs) for speaker age estimation by optimizing the mixture of classification and regression losses, achieving better performance than solo classification or regression based methods.
Recently, there is a line of work that utilize multi-task learning to estimate age and another speaker attribute, such as gender \cite{kwasny2021explaining}, and emotion \cite{si2021cross}.
 \cite{pan2021multi} applied a multi-task learning for the joint estimation of age and the Mini-Mental Status Evaluation criteria, showing improved performance on input features including i-vector and x-vector.



Although existing methods have achieved good performance on speaker age estimation, they ignored the fact that speaker age labels form a well-ordered numerical set and the age and the same speaker of adjacent ages produce speech signals hard to distinguish. 
The serious ordinal relationship and ambiguity between the labels should be well exploited to achieve improved accuracy.
In computer vision area, impressive progress on facial age estimation has been made, where label distribution learning (LDL) has shown great promise \cite{zhang2021practical}.



LDL \cite{wen2020adaptive} casts a classification problem into a distribution learning task by minimizing the discrepancy between predictions and constructed Gaussian distributions of labels. Gao et al. \cite{gao2017deep} used KL divergence to measure the similarity between the estimated and ground-truth distributions. And Pan et al. \cite{pan2018mean} proposed a multi-task approach while Wang et al. \cite{wang2019classification} introduced large margin classifier to it. 
In this work, we aim to propose a framework for speaker age estimation, which takes the audio features such as MFCC and i-vectors as input and produces precise age estimation with LDL method. Our framework naturally combines the classification and regression losses. Additionally, we regularize the variance of output age distribution to reduce the uncertainty of estimation.

The main contributions of this paper are summarized as follows: 
\begin{itemize}
    \item We propose a speaker age estimation framework with LDL method, which combines the regression and classification losses and a variance regularizer. This framework is applicable to many kinds of audio features, including MFCC, x-vector, i-vector, etc.
    \item Extensive experimental results on the public NIST SRE08-10 dataset and a real-world dataset illustrate that our LDL framework outperforms baseline methods by a significant margin.

\end{itemize}

\section{Proposed Method}
\label{sec:method}

\begin{figure*}[t]
\centering
\includegraphics[width=\textwidth]{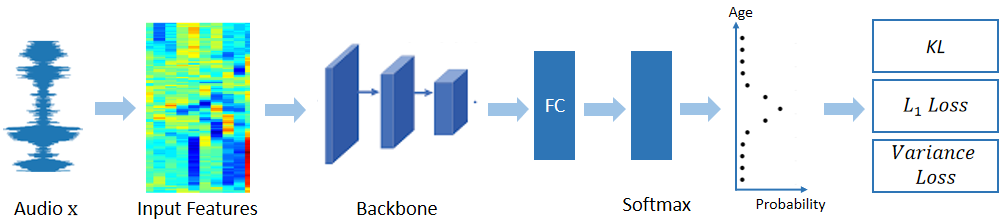}
\caption{ Overview of our proposed framework for speaker age estimation. For a given input utterance, we obtain its audio features (such as MFCCs), then feed the features through
a deep neural networks backbone to get high-level features, followed by a fully connected layer, which followed by both the regression and label distribution matching plus a variance regularizer.
}
\label{fig:arch}
\end{figure*}

\subsection{Network Architecture}

The pipeline of the proposed method has been outlined in Fig. \ref{fig:arch}, utilizing a DNN backbone to extract the embedding. The network backbone could be ResNet-18 \cite{he2016deep} or x-vector system \cite{snyder2017deep}, etc. The backbone yields embeddings of each speaker, which is further fed into a fully connected layer and finally a Softmax layer to get the output distribution of labels. 
Then for each sample, we compute three losses: 1.) the KL divergence to measure the discrepancy between the predicted label distribution and the discretized ground-truth one; 2.) the $\ell_1$ loss between the predicted age and the ground-truth age; and 3.) the variance of output age distribution. The predicted age is calculated based on the probability of every label in the Softmax layer.

\subsection{Label Distribution Learning}

Prior to the description of LDL, we make clear some notations. We use $k$ to denote the age label, $k\in[1, K]$ with $K$ the maximum age.
We use $n\in[1, N]$ to index a certain data sample, and $\hat{y}_{n}^{k}$ and $y_{n}^{k}$ are the predicted and ground-truth probabilities that the speaker age of $n$-th sample is $k$, respectively. $t_n$ is the ground-truth age for the $n$-th sample. We denote the mean of estimated age for the $n$-th sample:
\begin{equation}\label{eq:mean.age}
    \hat{y}_{n} = \sum_{k=1}^{K}k\cdot\hat{y}_{n}^{k}.
\end{equation}

The label distribution $\hat{y}$ is a probability distribution, which satisfy $\hat{y}_n^k \in [0, 1]$ and $\sum^{K}_{k=1}\hat{y}_n^k=1$. 
Typically, LDL proceeds by constructing the target age label distribution with a Gaussian distribution concentrating around the ground-truth age $t_n$. That is, for the data sample $n$, the probability ${y}_n^k$ is generated by the probability density function (p.d.f.) of a Gaussian distribution:
\begin{equation}
\begin{aligned}
{y}_n^k = \frac{1}{C_{n}}e^{{{ - \left( {k - t_n } \right)^2 } \mathord{\left/ {\vphantom {{ - \left( {x - \mu } \right)^2 } {2\sigma ^2 }}} \right. \kern-\nulldelimiterspace} {2\sigma ^2 }}},
\end{aligned}
\end{equation}
where $\sigma$ is a tuning hyper-parameter and $C_{n}$ is the normalizing constant to ensure $\sum_{k}{y}_n^k = 1$.

We employ the Kullback-Leibler (KL) divergence as the measurement of the discrepancy between ground-truth label distribution \textbf{\textit{${y}$}} and the prediction one $\hat{y}$.

Thus, we can define the KL loss function as follows:
\begin{equation}
L_{KL}(y,\hat{y})=\sum_{n=1}^{N}{D_{KL}(y_{n} | \hat{y}_{n})}
=\sum_{n=1}^{N}\sum_{k=1}^{K}{y_n^k log(\frac{y_n^k}{\hat{y}_n^k})},
\end{equation}
where $y_n=(y_n^1, \ldots, y_n^k)$ and $\hat{y}_n=(\hat{y}_n^1, \ldots, \hat{y}_n^k)$ are the ground-truth and predicted distributions of the $n$-th sample. When $\sigma$ is small enough, the ground-truth $y_{n}$ approaches to a categorical distribution, and $L_{KL}$ is equal to the cross-entropy loss of classification.

\subsection{Hybrid Loss}

Note that for each sample $n$, the LDL module learns a label distribution $\hat{y}_{n}^k$, so we can compute the mean and variance of the distribution. We apply $L_1$ loss to minimize the error between the predicted mean age $\hat{y}_n$ and ground-truth age ${y}_n$. $L_1$ is a regression loss.

\begin{equation}
\begin{aligned}
L_1(y,\hat{y})=\sum_N||y_n-\hat{y}_n||_1
\end{aligned}
\end{equation}

Our approach implicitly assumes that human voice aging process is episodic, implying that although voice aging is a continuous process, voice related to nearby ages are more related than far away ones. The variance loss penalizes the dispersion of estimated age distribution,

\begin{equation}
\begin{aligned}
L_v=\sum_N{\sum_K{\hat{y}_n^k * (k - \sum_{k=1}^{K} k \cdot \hat{y}_n^k)^2}}
\end{aligned}
\end{equation}

\noindent The overall loss is that given by, 

\begin{equation}
Loss =\lambda_1*L_{KL} + \lambda_2*L_1 + \lambda_3*L_v
\end{equation}
where $\lambda_1$, $\lambda_2$ and $\lambda_3$ are hyper-parameters.

\subsection{Inference}

In the inference phase, the age of a utterance is estimated as the mean age of predicted distribution in Eq. \eqref{eq:mean.age}.
Specifically, utterances are randomly cropped to 3-second clips during evaluation, we sum up the weighted estimation of all clips.

\section{Experiments}
\subsection{Datasets}
\textbf{NIST SRE08-10 dataset}
We conduct experiments on the public 2008-2010 NIST speaker recognition evaluation (SRE08-10) databases with configuration similar to \cite{Sadjadi_2016}. NIST SRE08 consists of 11205 utterances corresponding to 1227 speakers (769 female and 458 male) and SRE10 telephone condition consists of 5331 utterances corresponding to 492 speakers (256 female and 236 male). The dataset contains conversations in both English and non-English languages. Speech recordings from the short2-short3 core condition in the NIST SRE 2008 data are utilized for training the models, while speech
data from the NIST SRE 2010 telephony core condition are used as test material. There is no overlap between speech recordings extracted from the NIST 2008 and NIST 2010 SRE corpora (neither speakers nor recordings). 

\begin{figure}[h]
	\centerline{\includegraphics[width=.5\textwidth]{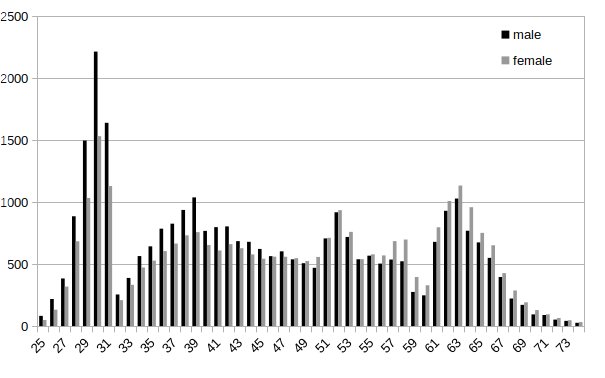}}
	\caption{The distribution of age and gender in PA-Age Dataset.}
	\label{fig:pa_age}
\end{figure}

\noindent\textbf{Real-world PA-Age Dataset}
The PA-Age dataset is collected from an large insurance company, which consists of 69610 utterances ranging from 10 second to 30 second corresponding to 59047 speakers (28386 female and 30661 male). The language of conversations are in Chinese and Chinese dialects. Utterances from the dataset are splitted randomly following the evaluation protocols of Subject-Exclusive (SE), where identities are randomly splitted into either train set or test set, but not both, to avoid label leakage. The test set is about 4000 utterances, and the rest utterances are used as training set. The age distribution of male and female is shown in Fig. \ref{fig:pa_age}. 


\subsection{Implementation Details}

\noindent \textbf{Baseline Methods:} As we propose a LDL-based framework for speaker age estimation, we compare our method with solo classification (by setting $\lambda_2=\lambda_{3}=0$ and $\lambda_2=1, \sigma=0.1$), solo regression (by setting $\lambda_1=\lambda_{3}=0$, $\lambda_2=1$), and mixture of classification and regression methods \cite{Ghahremani_2018} (by setting $\lambda_3 = 0$). Because our framework is compatible to many speech feature extractors, we also conduct experiments using i-vector, x-vector, and ResNet-18 backbone.

For all three speaker embeddings, the speech features were 23-dim MFCC short-time mean normalized over sliding the window of 3 seconds. An energy-based SAD
(speech activity detection) was used to remove non-speech frames.

\noindent\textbf{i-Vector system:} The MFCC features with short-time centering were used as input to the GMM-UBM model \cite{reynolds2000speaker}.
The UBM and i-vector extractors were trained on
NIST SRE04-06 English telephone speech containing 1936 female speakers and 679 male speakers. We used a 2048 component GMM-UBM model with full covariance matrices. Total
variability subspace dimension was set to 400. It is worth mentioning that there is no speaker overlap between the data used
to train the i-vector extractor and data used to train and test the
age estimation LDL/ResNet-18/x-vector system. The input dimension for the fully-connected layer in Fig. \ref{fig:arch} was 400, and only one hidden layer with 256 neurons and  rectified linear
unit (ReLU) activation function.

\noindent\textbf{x-Vector system:} The MFCC features with short-time centering were used as input to the x-vector architecture. The
time-delay deep network layer (TDNN) with the ReLU non-linearity was used. Batch normalization was
 also used after the non-linearity. Details can be found in \cite{Ghahremani_2018}.

\noindent\textbf{ResNet-18 system:}
ResNet-18 consists of 18 residual blocks
stacked on top of each other. The residual block has
two $3\times3$ convolutional layers with the same number of
output channels. Each convolutional layer is followed
by a batch normalization layer and a ReLU activation
function. A skip connection is added which skips these
two convolution operations and adds the input directly
before the final ReLU activation function. The objective
of the skip connections is to perform identity mapping.



All experiments are conducted using the PyTorch and ASVTorch \cite{lee2021asvtorch} framework. The mini-batch size was set to 32. We used stochastic gradient descent (SGD) optimizer with an initial
learning rate of 0.001 and a momentum of 0.9. We decreased the learning rate by a factor of 2 when the validation loss does
not improve for two successive epochs. Minimum learning rate was set to 1e-5.

\subsection{Metrics}

To assess the goodness of our age estimators, we report performance in terms of mean absolute error (MAE) and Pearson’s correlation
coefficient. MAE is defined,
$$
\text{MAE}=\frac{1}{N}\sum^{N}_{i=1}|y_{i}-\textit{t}_{i}|
$$
where $\textit{t}_{i}$ and $y_{i}$ are the ground-truth and estimated age of the $i$-th example, respectively.
Pearson’s correlation coefficient is defined as,
$$
\rho = \frac{1}{N-1}\sum_{i=1}^{N}\Big(\frac{y_{i}-\mu_{y}}{\sigma_{y}}\Big)\Big(\frac{t_{i}-\mu_{t}}{\sigma_{t}}\Big)
$$
where $\mu_y$ and $\sigma_y$ are the mean and standard deviation for the
predicted ages; and $\mu_t$ and $\sigma_t$ for true ages. Higher correlation coefficients are better.

\subsection{Results and Analyses}

We compare our framework with baseline methods commonly used in existing literature. The results on two datasets are reported in Table \ref{tab:main.res}. In this table, Reg, Cls and Reg+Cls represent solo regression ($\lambda_1=\lambda_3=0, \lambda_2=\sigma=1$), solo classification ($\lambda_2=\lambda_3=0, \lambda_1=1, \sigma=0.1$) and mixture of regression and classification ($\lambda_3=0, \lambda_1=\lambda_2=1, \sigma=0.1$), respectively. All models are trained on 5 seconds speech trunks. 
From this table, for both datasets and all three feature extractors (i-vector, x-vector and ResNet-18), our LDL-based framework ($\lambda_1=\lambda_2=1, \lambda_3=0.1, \sigma=1$) significantly outperforms the baseline methods in terms of both MAE and Person's correlation. The regression method perform the worst, followed by classification, which is consistent to findings in the literature. In terms of feature extraction, ResNet-18 performs the best, followed by x-vector. For the regression, i-vector outperforms x-vector, which may be caused by the poor trained x-vector architecture. In general, our framework produces more than 10\% reduction in MAE on the real-world PA-Age dataset.


\begin{table}[h]
	\centering
	\caption{Comparison of different methods and feature extractors on SRE08-10 and PA-Age Datasets, in which Reg, Cls and Reg+Cls represent solo regression, solo classification and mixture of regression and classification, respectively.\label{tab:main.res}}\smallskip
	\resizebox{0.48\textwidth}{!}{
\begin{tabular}{lllllll}
\hline
\multirow{2}{*}{SRE08-10} & \multicolumn{2}{l}{i-Vector} & \multicolumn{2}{l}{x-Vector} & \multicolumn{2}{l}{ResNet-18} \\
                          & MAE           & $\rho$          & MAE           & $\rho$          & MAE            & $\rho$          \\ \hline
Reg                       & 8.54          & 0.71         & 9.12          & 0.70         & 7.96           & 0.72         \\
Cls                       & 6.18          & 0.78         & 6.03          & 0.78         & 5.52           & 0.80         \\
Reg+Cls                   & 5.54          & 0.80         & 5.25          & 0.81         & 5.14           & 0.82         \\ \hline
\textbf{LDL(Ours)}                 & \textbf{4.98}          & \textbf{0.85}         & \textbf{4.75}          & \textbf{0.86}         & \textbf{4.62}           & \textbf{0.87}         \\ \hline
PA-Age                    &               &              &               &              &                &              \\ \hline
Reg                       & 11.69         & 0.61         & 12.21         & 0.60         & 10.92          & 0.63         \\
Cls                       & 8.75          & 0.69         & 8.37          & 0.71         & 8.20           & 0.72         \\
Reg+Cls                   & 7.89          & 0.74         & 7.45          & 0.75         & 7.03           & 0.76         \\ \hline
\textbf{LDL(Ours)}                 & \textbf{6.97}          & \textbf{0.78}         & \textbf{6.34}          & \textbf{0.81}         & \textbf{6.23}           & \textbf{0.82}         \\ \hline
\end{tabular}
}
\end{table}

Table \ref{tab:ablation} displays the ablation study of hyper-parameters $\lambda_3$ and $\sigma$ by setting $\lambda_1 = \lambda_2 = 1$ and using ResNet-18 as the backbone. In this table, we evaluate the MAE of our LDL framework at different combination of values of $\lambda_3$ and $\sigma$. When $\lambda_3$ is fixed, increasing $\sigma$ from 0.5 to 1.0, the performance of LDL method is increasing in terms of MAE. The best performance is achieved at $\lambda_3 = 0.1$ and $\sigma=1.0$.

\begin{table}[h]
	\centering
	\caption{Ablation study of $\lambda_3$ and $\sigma$ on the PA-Age dataset by setting $\lambda_1 = \lambda_2 =1$ and using ResNet-18 backbone. }\smallskip
	\resizebox{0.48\textwidth}{!}{
		\begin{tabular}{ccccccc}
			\hline
		$\lambda_3$ & 0.01 & 0.1 & 0.1 & 1.0 & 1.0 & 10.0 \\
		$\sigma$  &0.1 & 0.5 & 1.0 &0.5 &1.0 & 3.0 \\\hline
		\textbf{LDL(ours)}    & 8.52& 8.15 &\textbf{6.23}& 7.89 & 7.45 & 8.37\\ \hline
		\end{tabular}
	}
	\label{tab:ablation}
\end{table}

Table \ref{tab:seg.length}
shows the effects of different training/test durations over LDL with ResNet-18 backbone. With the increase of test segment length, the MAE drops significantly. As the training segments increases, the test performance drops because of overfitting.
\vspace{-5mm}
\begin{table}[h]
	\centering
	\caption{Effect of training/test segment length for age estimation on PA-Age using ResNet-18 backbone.}\smallskip
	\resizebox{0.48\textwidth}{!}{
\begin{tabular}{lllll}
\hline
MAE                     & \multicolumn{4}{l}{Test segment length(s)} \\
Train segment length(s) & 10        & 15        & 20       & full    \\ \hline
5                       & \textbf{13.16}     & \textbf{11.04}     & \textbf{6.14}     & \textbf{6.10}    \\
10                      & 15.31     & 12.96     & 6.25     & 6.17    \\ \hline
\end{tabular}
}
\label{tab:seg.length}
\end{table}

\section{Conclusions}
\label{sec:print}
In this paper, we propose a LDL-based framework for speaker age estimation, which combines the regression and classification objectives and a variance minimization. Our experiments verify its effectiveness on both public and real-world datasets, so it could have good potential in many real applications.


\section{Acknowledgment}
This paper is supported by the Key Research and Development Program of Guangdong Province under grant No. 2021B0101400003 and the National Key Research and Development Program of China under grant No. 2018YFB0204403. Corresponding author is Jianzong Wang from Ping An Technology (Shenzhen) Co., Ltd (jzwang@188.com).

\newpage
\bibliographystyle{IEEEbib}
\bibliography{strings,refs}

\end{document}